\documentclass[backref, breaklinks, twocolumn, colorlinks]{aastex631}   
\hypersetup{linkcolor=blue,citecolor=blue,filecolor=cyan,urlcolor=black}
 
\usepackage{latexsym,amsmath,amssymb}
\usepackage{verbatim}
\usepackage{multirow}
\usepackage{color}
\usepackage[hang]{footmisc}
\usepackage{soul,xcolor}
\usepackage{afterpage}

\bibpunct{(}{)}{;}{a}{}{,}

\newcommand {\xmm} {\textsl{XMM-Newton}}

\newcommand {\swift} {\textsl{Swift}}

\newcommand {\nicer} {\textsl{NICER}}

\def \rsun {\ifmmode$R$_{\odot}\else R$_{\odot}$}

          \font\sixrm=cmr6

\def\rns{R_{\hbox{\sixrm NS}}}

\def\teq#1{$\, #1\,$}                         

\def \hcm {\hbox {\ifmmode $ atoms cm$^{-2}\else atoms cm$^{-2}$\fi}}

\def\approxgt{\mathrel{\hbox{\rlap{\lower.55ex \hbox {$\sim$}}
        \kern-.3em \raise.4ex \hbox{$>$}}}}
\def\approxlt{\mathrel{\hbox{\rlap{\lower.55ex \hbox {$\sim$}}
        \kern-.3em \raise.4ex \hbox{$<$}}}}

\newcommand\T{\rule{0pt}{2.6ex}}
\newcommand\B{\rule[-1.2ex]{0pt}{0pt}}

\def \src {SGR 1830$-$0645}

\begin{document}
\setstcolor{red}

\title{\rm \uppercase{Pulse Peak Migration during the Outburst Decay
    of the Magnetar \src:\\ Crustal Motion and Magnetospheric Untwisting}}

%
\author[0000-0002-7991-028X]{George~Younes}
\affiliation{Astrophysics Science Division, NASA Goddard Space Flight Center, Greenbelt, MD 20771, USA}
\affiliation{Universities Space Research Association (USRA) Columbia, Maryland 21046, USA}

\author{Samuel~K~Lander}
\affiliation{Physics, University of East Anglia, Norwich, NR4 7TJ, U.K.}

\author[0000-0003-4433-1365]{Matthew~G.~Baring}
\affiliation{Department of Physics and Astronomy - MS 108, Rice University, 6100 Main Street, Houston, Texas 77251-1892, USA}

\author[0000-0003-1244-3100]{Teruaki Enoto}
\affiliation{RIKEN Cluster for Pioneering Research, 2-1 Hirosawa, Wako, Saitama 351-0198, Japan}

\author[0000-0003-1443-593X]{Chryssa~Kouveliotou}
\affiliation{Department of Physics, The George Washington University, Washington, DC 20052, USA, gyounes@gwu.edu}
\affiliation{Astronomy, Physics and Statistics Institute of Sciences (APSIS), The George Washington University, Washington, DC 20052, USA}

\author[0000-0002-9249-0515]{Zorawar Wadiasingh}
\affiliation{Department of Astronomy, University of Maryland, College Park, Maryland 20742, USA}
\affiliation{Astrophysics Science Division, NASA Goddard Space Flight Center, Greenbelt, MD 20771, USA}
\affiliation{Center for Research and Exploration in Space Science and Technology, NASA/GSFC, Greenbelt, Maryland 20771, USA}

\author[0000-0002-6089-6836]{Wynn C. G. Ho}
\affiliation{Department of Physics and Astronomy, Haverford College, 370 Lancaster Avenue, Haverford, PA 19041, USA}

\author{Alice~K.~Harding}
\affiliation{Theoretical Division, Los Alamos National Laboratory, Los Alamos, NM 87545, USA}

\author{Zaven Arzoumanian}
\affiliation{Astrophysics Science Division, NASA Goddard Space Flight Center, Greenbelt, Maryland 20771, USA}

\author[0000-0001-7115-2819]{Keith Gendreau}
\affiliation{Astrophysics Science Division, NASA Goddard Space Flight Center, Greenbelt, Maryland 20771, USA}

\author[0000-0002-3531-9842]{Tolga G\"uver}
\affiliation{Istanbul University, Science Faculty, Department of Astronomy and Space Sciences, Beyaz\i t, 34119, Istanbul, Turkey}
\affiliation{Istanbul University Observatory Research and Application Center, Istanbul University 34119, Istanbul Turkey}

\author[0000-0001-8551-2002]{Chin-Ping Hu}
\affiliation{Department of Physics, National Changhua University of Education, Changhua 50007, Taiwan}

\author[0000-0002-0380-0041]{Christian~Malacaria}
\affiliation{NASA Marshall Space Flight Center, NSSTC, 320 Sparkman Drive, Huntsville, AL 35805, USA}
\affiliation{Universities Space Research Association, Science and Technology Institute, 320 Sparkman Drive, Huntsville, AL 35805, USA}

\author[0000-0002-5297-5278]{Paul S. Ray}
\affiliation{Space Science Division, U.S. Naval Research Laboratory, Washington, DC 20375, USA}

\author[0000-0001-7681-5845]{Tod E. Strohmayer}
\affiliation{Astrophysics Science Division and Joint Space-Science Institute, NASA's Goddard Space Flight Center, Greenbelt, MD 20771, USA}

\begin{abstract}

Magnetars, isolated neutron stars with magnetic field strengths
typically $\gtrsim10^{14}$~G, exhibit distinctive months-long outburst
epochs during which strong evolution of soft X-ray pulse profiles,
along with nonthermal magnetospheric emission components, is often
observed. Using near-daily \nicer\ observations of the magnetar
SGR~1830-0645 during the first 37 days of a recent outburst decay, a
pulse peak migration in phase is clearly observed, transforming the
pulse shape from an initially triple-peaked to a single-peaked
profile. Such peak merging has not been seen before for a
magnetar. Our high-resolution phase-resolved spectroscopic analysis
reveals no significant evolution of temperature despite the complex
initial pulse shape. Yet the inferred surface hot spots shrink during
the peak migration and outburst decay. We suggest two possible origins
for this evolution. For internal heating of the surface, tectonic
motion of the crust may be its underlying cause. The inferred speed of
this crustal motion is $\lesssim100$~m~day$^{-1}$, constraining the
density of the driving region to $\rho\sim10^{10}$~g~cm$^{-3}$, at a
depth of $\sim200$~m. Alternatively, the hot spots could be heated by
particle bombardment from a twisted magnetosphere possessing flux
tubes or ropes, somewhat resembling solar coronal loops, that untwist
and dissipate on the 30-40~day timescale. The peak migration may then
be due to a combination of field-line footpoint motion (necessarily
driven by crustal motion) and evolving surface radiation
beaming. These novel dataset paints a vivid picture of the dynamics
associated with magnetar outbursts, yet it also highlights the need
for a more generic theoretical picture where magnetosphere and crust
are considered in tandem.

\end{abstract}

\section{Introduction}
\label{Intro}

Magnetar outburst epochs start with an increase of the quiescent X-ray flux by as many as three orders of magnitude, accompanied by drastic spectral changes to the neutron star's surface thermal and magnetospheric emission, as well as strong temporal variability in the form of timing noise, glitch activity, and altered pulse shape \citep[e.g.,][]{woods04ApJ:1E2259,gavriil06ApJ:1048,rea09MNRAS:sgr0501,israel10MNRAS:1547,esposito10MNRAS:0418,gavriil11ApJ:0142,scholz14ApJ:1822,hu20ApJ:1818}. The outbursts last anywhere from months to years during which the source properties typically return back to their initial state \cite[][but also see \citealt{younes17ApJ:1806,cotizelati2020AA:1547}]{cotizelati18MNRAS}. Given these variability patterns, outburst epochs are thus distinctly revealing of a magnetar's highly dynamic magnetosphere and its interplay with the surface thermal emission, both of which are governed by the decay of the super-critical internal and external B-fields \citep[e.g.,][]{thompson95MNRAS:GF,thompson96ApJ:magnetar,vigano13MNRAS}.

\src\ was discovered on 2020 October 10 after a short, soft burst from
its direction was detected with the  \swift/BAT instrument
\citep{page20ATel14083}. Subsequent dedicated X-ray campaigns revealed
the presence of a bright X-ray source with rotational properties
consistent with the bulk of the magnetar family; a rotational
frequency $\nu=0.096$~Hz undergoing a spin down at a rate of
$\dot\nu=-6.2\times10^{-14}$~Hz~s$^{-1}$ \citep[][Younes et al. 2021,
accepted, hereinafter Y21]{cotizelati21ApJ1830}, implying a dipole
field strength $B=2.7\times10^{14}$~G at the equator and a spin-down
age of 24.3~kyr. In the several months following the source discovery,
its soft X-ray flux decreased by a factor 6 due to the shrinkage of
the total emitting area (Y21)

In this paper, we present a study of the thermal pulse shape temporal evolution of \src\ as observed with \nicer. We also perform the most detailed phase-resolved spectroscopic analysis of the soft thermal surface emission of any magnetar to date. The following section summarizes the observations and data reduction. Section~\ref{res} describes the analysis and presents the results of our campaign. We conclude in Section~\ref{discuss} with a discussion on the implications of the surface heat-map of this magnetar as well as the role of the crust in triggering the outburst in \src\ and perhaps in the magnetar population as a whole.

\begin{figure}
    \centering
    \includegraphics[width=3.4in]{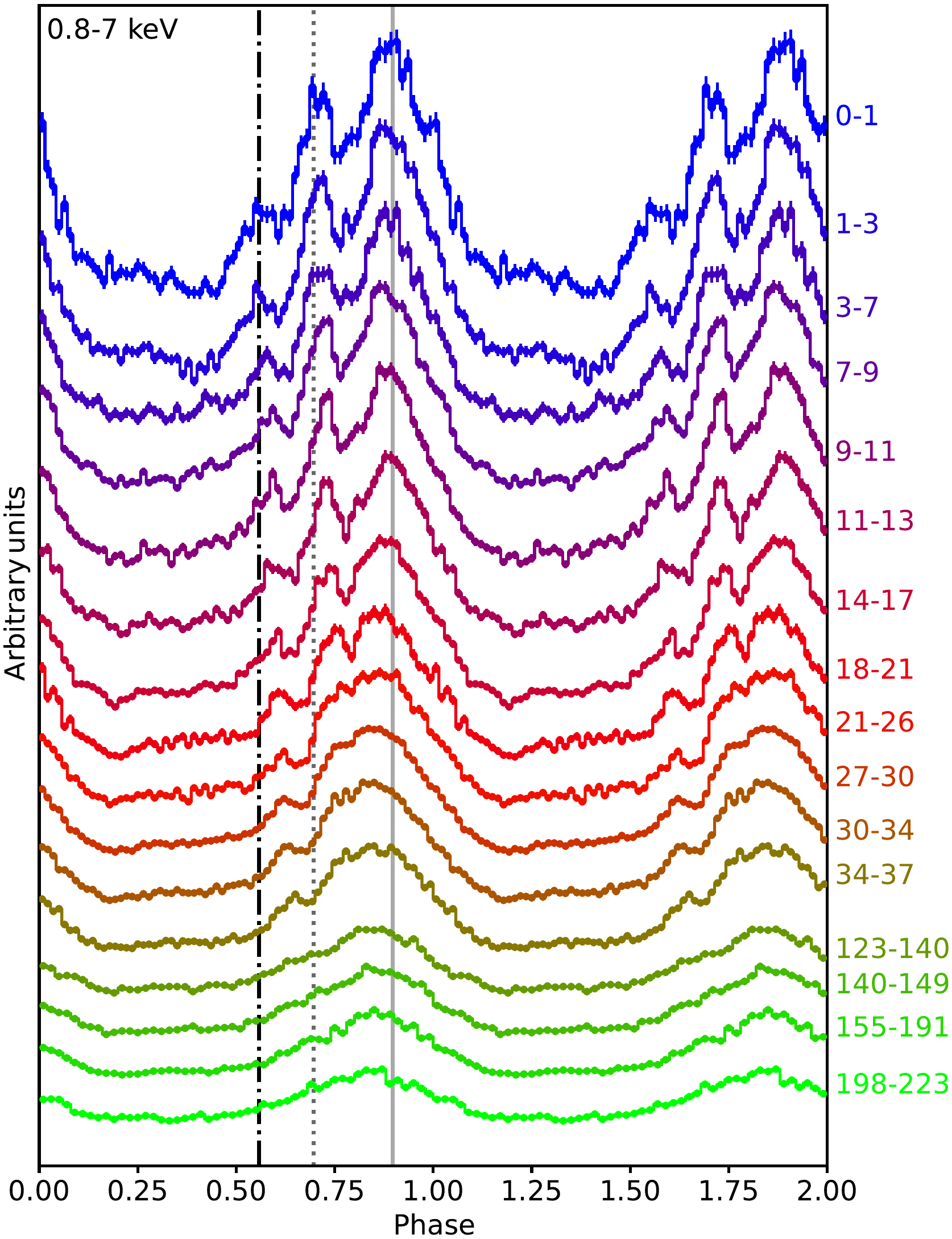}
    \caption{The 0.8--7~keV pulse profile evolution of \src\ with
      time. The black dot-dashed, dark-grey dotted, and light-grey
      solid lines represent the Gaussian centroids of the three peaks
      during the first day of outburst. A shift, towards the right for
      the two weaker peaks and the left for the brightest peak, is
      evident. This shift produces a simpler, nearly sinusoidal, pulse
      shape four months after outburst onset. The numbers to the right
      indicate the intervals, in days from outburst onset, that were
      used to derive each pulse. Adapted from Y21.}
    \label{fig:ppCount}
\end{figure}

\section{Observations and data reduction}
\label{obs}

\nicer\ observed \src\ with almost daily cadence since its discovery on 2020 October 10 and up to 2020 November 17, after which the source was sun-constrained (i.e., too close to the Sun) and could not be observed. \nicer\ restarted its monitoring campaign on 2021 February 10 with weekly observations. The details of these observations, their  temporal and phase-averaged spectroscopic analyses, the burst analysis, as well as radio non-detection limits of the source are detailed in Y21. In this Letter, we focus on the
phase-resolved spectroscopic analysis as well as pulse-shape temporal evolution during the first 37 days of the outburst. The near-daily
cadence has allowed us to track the temporal evolution of these two elements with unprecedented detail. We complement our analysis with
the data stretching from 2021 February to May for comparison purposes.

For any pulse and phase-resolved analysis we present in this Letter,
we utilize the phase-coherent timing solution presented in Y21. The
spectral analysis is performed using Xspec version 12.11.0m
\citep{arnaud96conf} in the energy range 0.8--7~keV. We group all
spectra to have 5 counts per bin. We determine the background spectra
for each observation using the {\texttt nibackgen3C50} method; we
added a conservative 20\% systematic uncertainty to the estimated
background number counts per \nicer\ energy channel
\citep{remillard2021:3c50}. We derive best-fit model parameters
and their associated uncertainties utilizing the \texttt{pgstat}
statistic, which is valid for the case of Poisson distributed data
with background having a Gaussian distributed uncertainties, as is the
case for \nicer. We quote all uncertainties at the $68\%$ confidence
level, unless otherwise noted.

\section{Results}
\label{res}

\begin{figure*}
    \centering
    \includegraphics[angle=0,width=0.48\textwidth]{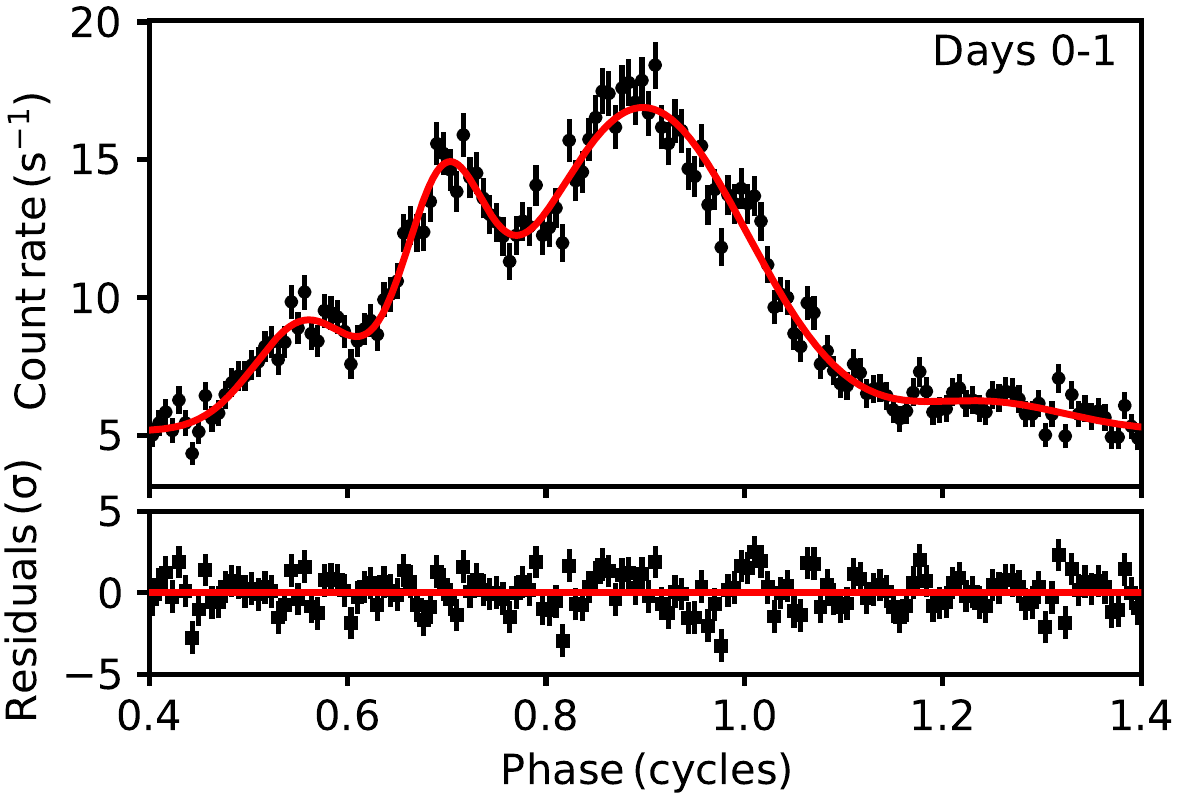}
    \includegraphics[angle=0,width=0.48\textwidth]{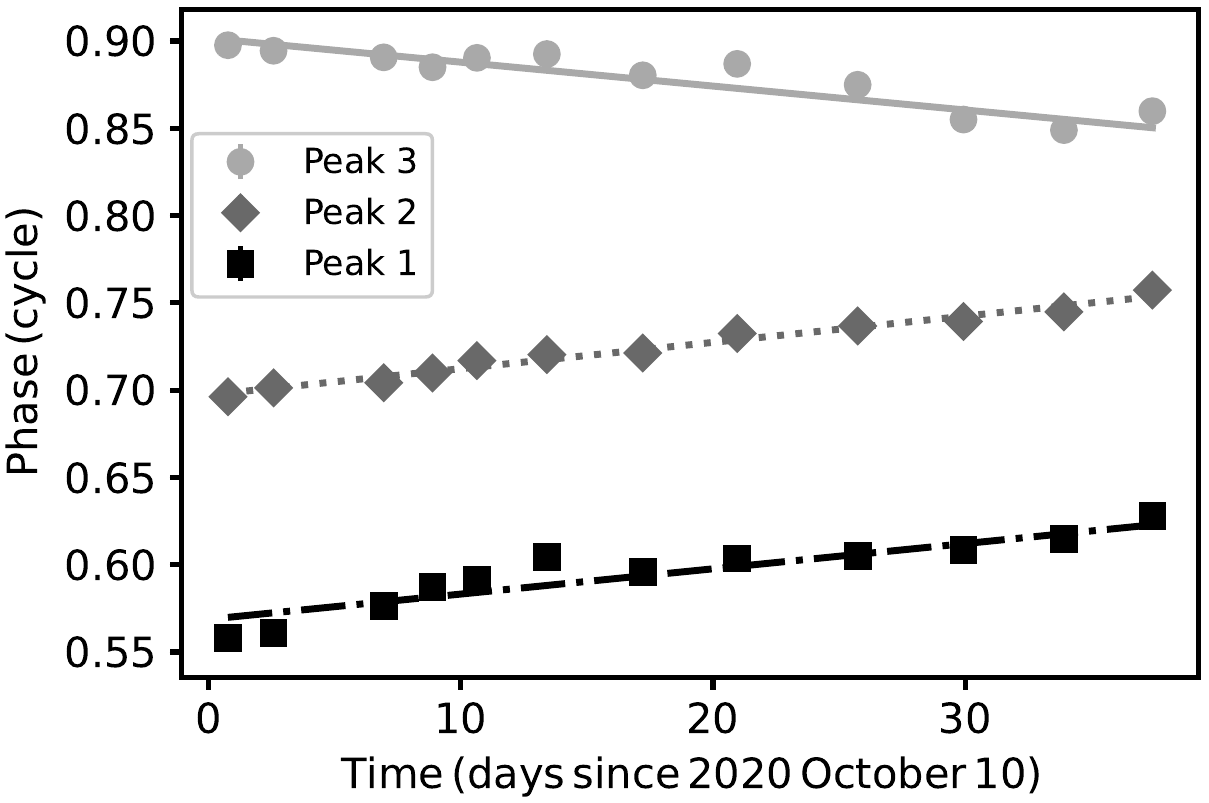}
    \caption{{\sl Upper-left panel.} The \src\ pulse profile, shown as black points with corresponding uncertainties as solid vertical bars, derived from the first day of \nicer\ observations at outburst onset (i.e., 2020 October 10). The red solid line represents the best-fit four-Gaussian model to the data. {\sl Lower-left panel.} Residuals from the best-fit model in terms of $\sigma$. {\sl Right panel.} Temporal evolution of the best-fit Gaussian centroids to the three peaks in the \src\ pulse profiles (the corresponding $68\%$ uncertainties are smaller than the symbol size). All three peaks follow a similar linear trend with an absolute rate of about $(1.5 \pm 0.1) \times 10^{-3}$~cycle~day$^{-1}$. Note the opposite motion of peak 3, the brightest, compared to the other two. See text for more details.}
    \label{fig:ppShift}
\end{figure*}

\subsection{Pulse-shape evolution}
\label{resPPE}

Figure~\ref{fig:ppCount} shows the 0.8-7~keV folded pulse profiles (two rotational cycles are plotted for clarity) at distinct epochs from source discovery, averaged over one to several days. At the early stages of the outburst, the profile is markedly triple-peaked, yet, with a clear phase shift in each pulse-peak (Y21). We resort to simple Gaussian fits to the pulse profiles to identify the centroid of each peak at different epochs. For this, we group each profile to have 150 phase bins, ensuring a minimum number of 50 counts in each. We utilize maximum-likelihood estimation to derive the best-fit model to the data, which were assumed to be Gaussian distributed. The goodness of fit is assessed from the $\chi^2$ statistics. We obtained $1\sigma$ uncertainties on the best-fit parameters by using the Markov Chain Monte Carlo sampler \texttt{emcee} \citep{Foreman13PASP:emcee} and assuming flat priors.  Finally, we model the pulse profiles with $\phi\in[0.4,1.4)$ to ensure that none of the apparent main pulses fall at the edge of our fitting parameter space.

We start by fitting the pulse profile averaged over the first day of
the outburst with an increasing number of Gaussian components (plus a
constant, non-pulsed component) up to the term deemed statistically
significant at the $3\sigma$ confidence level through an
F-test. We find that a model consisting of four Gaussians is
sufficient to describe the data with a reduced $\chi^2$ of 1.27 for
137 degrees of freedom (dof); adding a fifth does not improve the
quality of the fit. The left panel of Figure~\ref{fig:ppShift} shows
this best-fit model along with the residuals in terms of $\sigma$. The
three peaks are well sampled with the Gaussian curves, while a fourth,
low-amplitude one, is required to fit the off-pulse part of the
profile. In  Figure~\ref{fig:ppCount}, we display the Gaussian
centroids of these three main pulses, to which we refer, hereafter, as
peak 1, peak 2, and peak 3 in ascending phase order.

Subsequently, we fit the rest of the pulse profiles shown in Figure~\ref{fig:ppCount} up to day 37 with the same model, which resulted in a statistically acceptable fit for each case ($\chi^2_{\nu}\approx1$). The right panel of Figure~\ref{fig:ppShift} presents the temporal evolution of the Gaussian centroids for the three main peaks, along with their best-fit linear trends. We find that peak 1, peak 2, and peak 3 are shifting at a rate of $(1.5 \pm 0.1) \times 10^{-3}$, $(1.5 \pm 0.1) \times 10^{-3}$, and $(-1.4 \pm 0.1) \times 10^{-3}$~cycle~day$^{-1}$, implying that the shift is occurring to simplify the pulse shape via the merging of the different peaks.

This rate of motion projected on the sky translates to a speed $\sim100$~m~day$^{-1}$ assuming the motion is occurring along the orthodrome and a star radius of $10$~km. This estimate ignores gravitational light-bending, and other elements, such as reprocessing of the soft thermal emission in the highly magnetized atmosphere and the exact shape of the emitting regions. Fitting each profile with physically-motivated models is beyond the scope of the paper. Yet, the latter two elements will not have a strong impact on the centroid of each pulse peak but more so on their wings \citep{vanAdelsberg06MNRAS,taverna20MNRAS,Barchas-2021-MNRAS}, while gravitational light-bending tends to slightly increase the visible area over which the motion is occurring. Hence, this projected speed should be considered an order-of-magnitude estimate. We discuss these results in detail in Section~\ref{discuss}.

\subsection{Phase-resolved spectroscopy}
\label{resPRS}

We perform our phase-resolved spectroscopic analysis for each of the
pulse profiles presented in Figure~\ref{fig:ppCount}, grouped to 50
(20 for the post-Sun-constrained period) phase bins. This choice was
driven by the balance to increase S/N in each phase bin while
retaining the main features in the profiles. We fit all 50 bins of
each profile with an absorbed double-blackbody (BB) model, which
provided the best fit to the phase-averaged spectra of the source
throughout the outburst (Y21). We fix the absorption
column density to the value derived from the latter fit, $N_{\rm
  H}=1.17\times10^{22}$~cm$^{-2}$ (Y21). We start by
letting the temperatures ($kT$) and areas of the thermally emitting
regions vary freely. We find that the BB temperatures show no
dependence on phase within their uncertainties. We test the level of
scatter around the average by fitting a horizontal line to $kT$ versus
phase and measuring the reduced $\chi^2$, which we find to be in the
range of 0.9-1.5, implying little scatter. This temperature constancy
with phase was also noted during the \xmm\ observation of the source
obtained on October 11 and 12 \citep{cotizelati21ApJ1830}. Finally, we
find that the temperatures matched the values derived through the
phase-averaged spectroscopic analysis with \nicer\ ($kT_{\rm
  warm}\approx0.45$~keV and $kT_{\rm hot}\approx1.2$~keV), which also
reveals constant BB temperatures throughout the outburst (Y21). Hence, we fix the temperatures of each phase-bin spectral model to these latter measurements. As an extra layer of caution, we test the goodness of this fit to each phase bin by simulating 1000 spectra drawn from the best-fit model and noting the percentage of the test statistics of these simulated spectra that are smaller than the true one. This percentage is consistently around $50\%$, implying that this model provides an accurate representation of the data.

\begin{figure*}[t!]
\begin{center}
  \includegraphics[angle=0,width=0.49\textwidth]{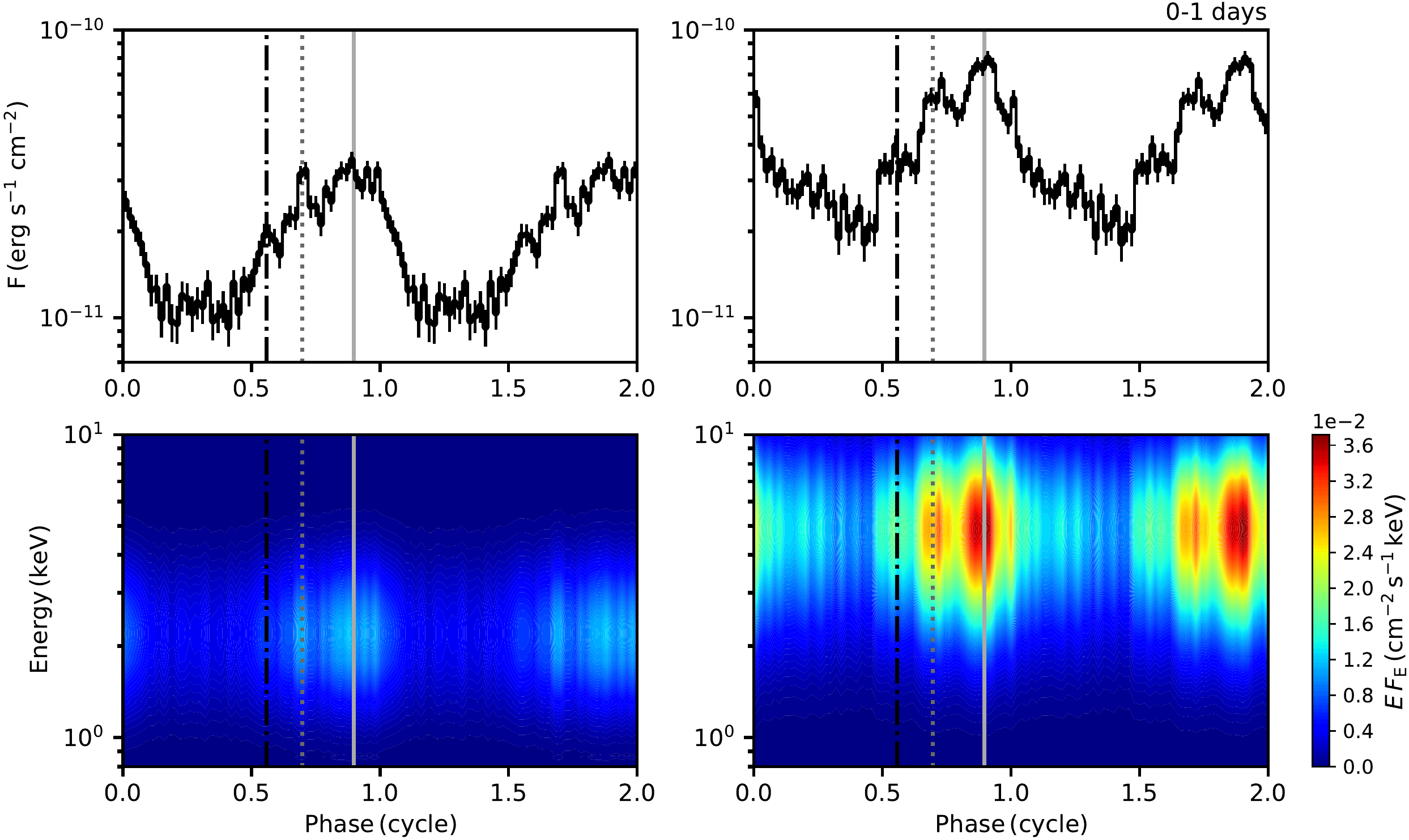}
  \includegraphics[angle=0,width=0.49\textwidth]{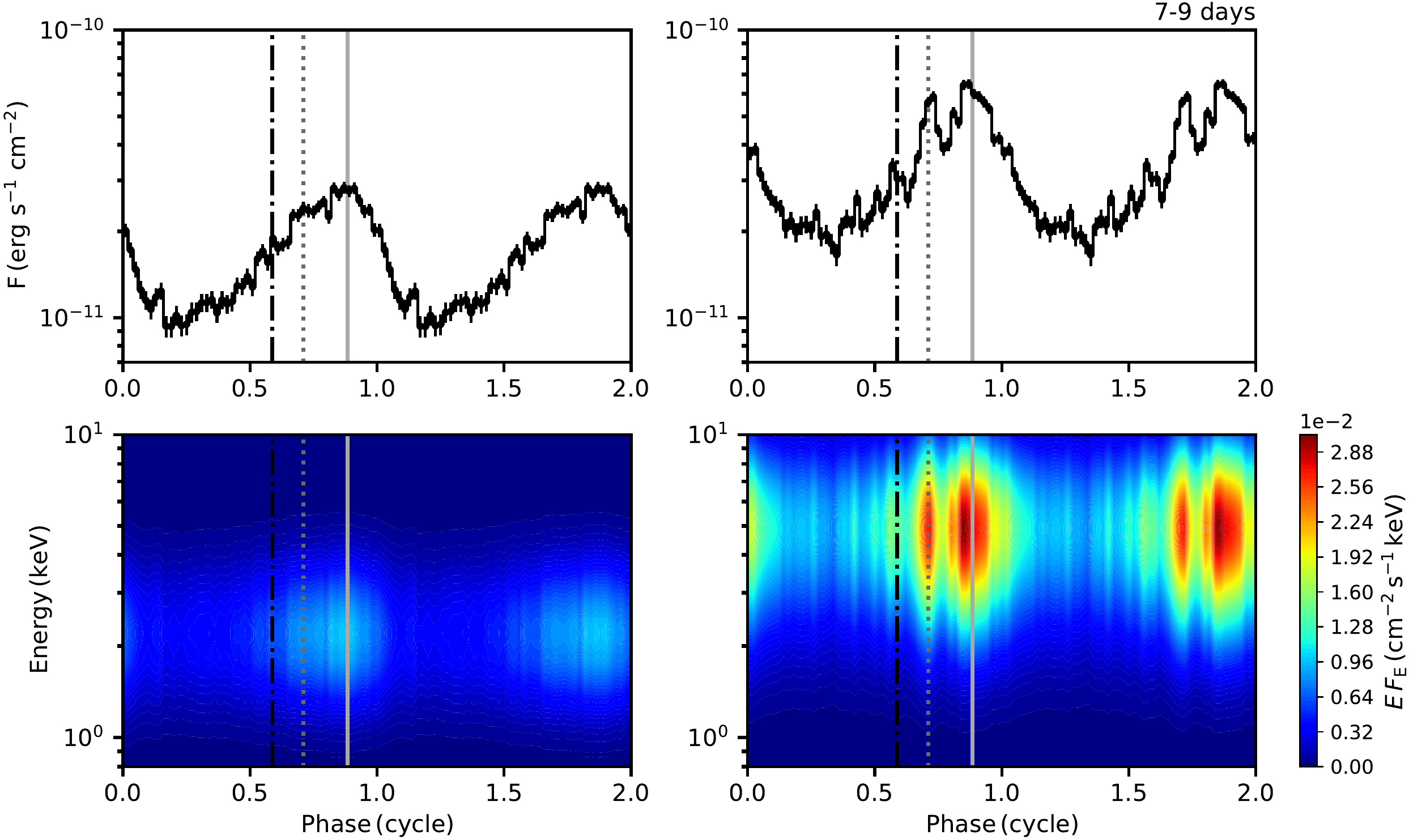}\\
  \includegraphics[angle=0,width=0.49\textwidth]{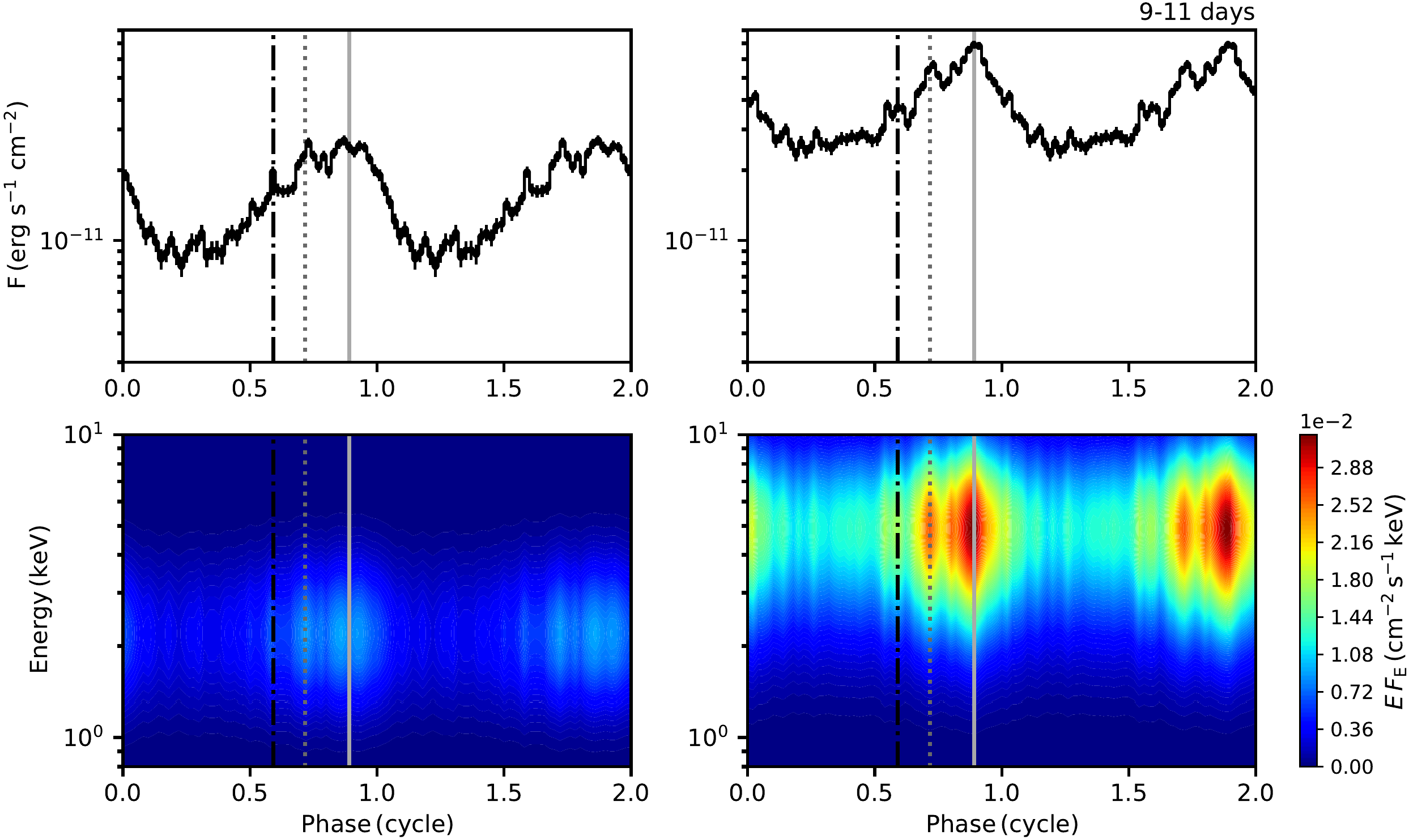}
  \includegraphics[angle=0,width=0.49\textwidth]{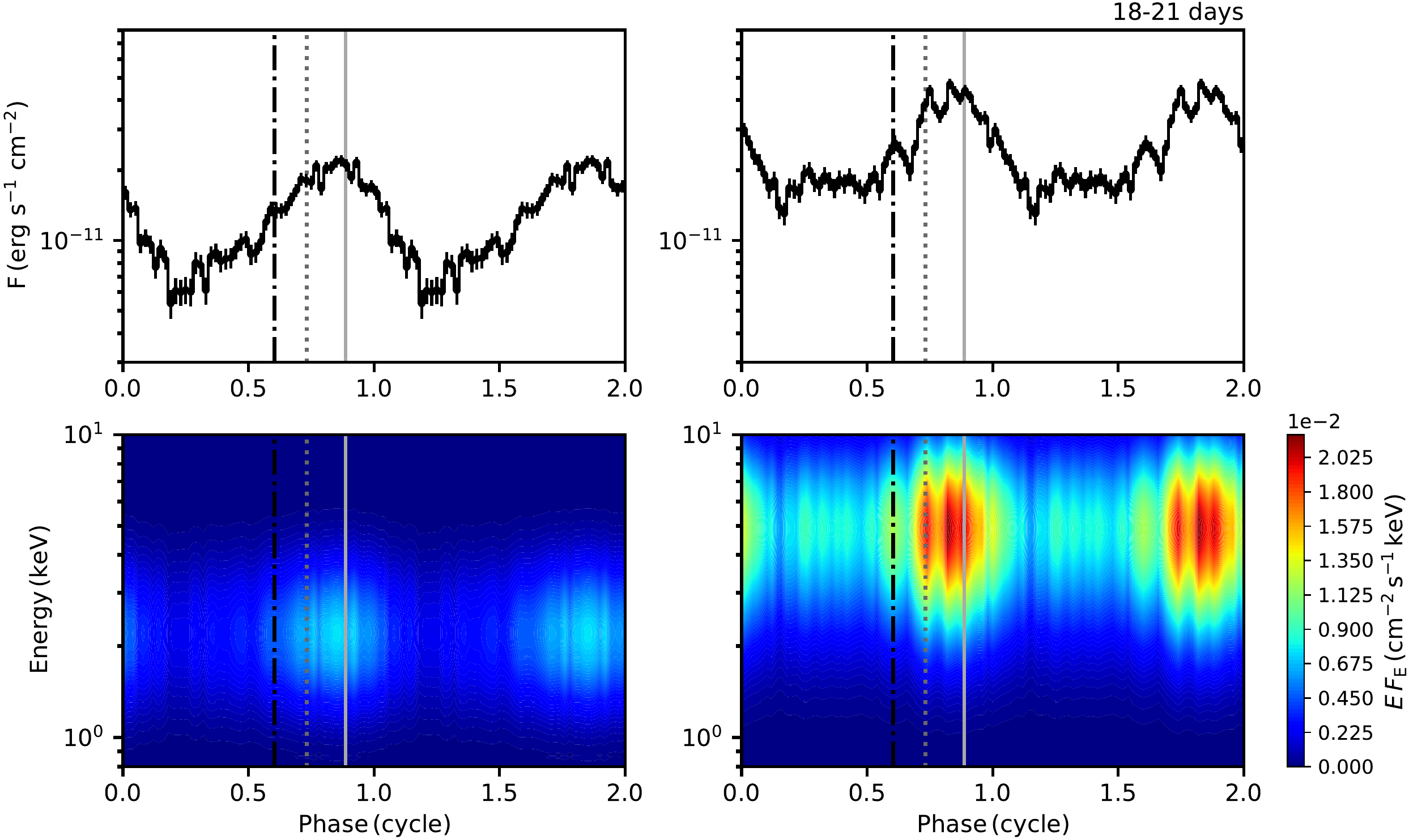}\\
  \includegraphics[angle=0,width=0.49\textwidth]{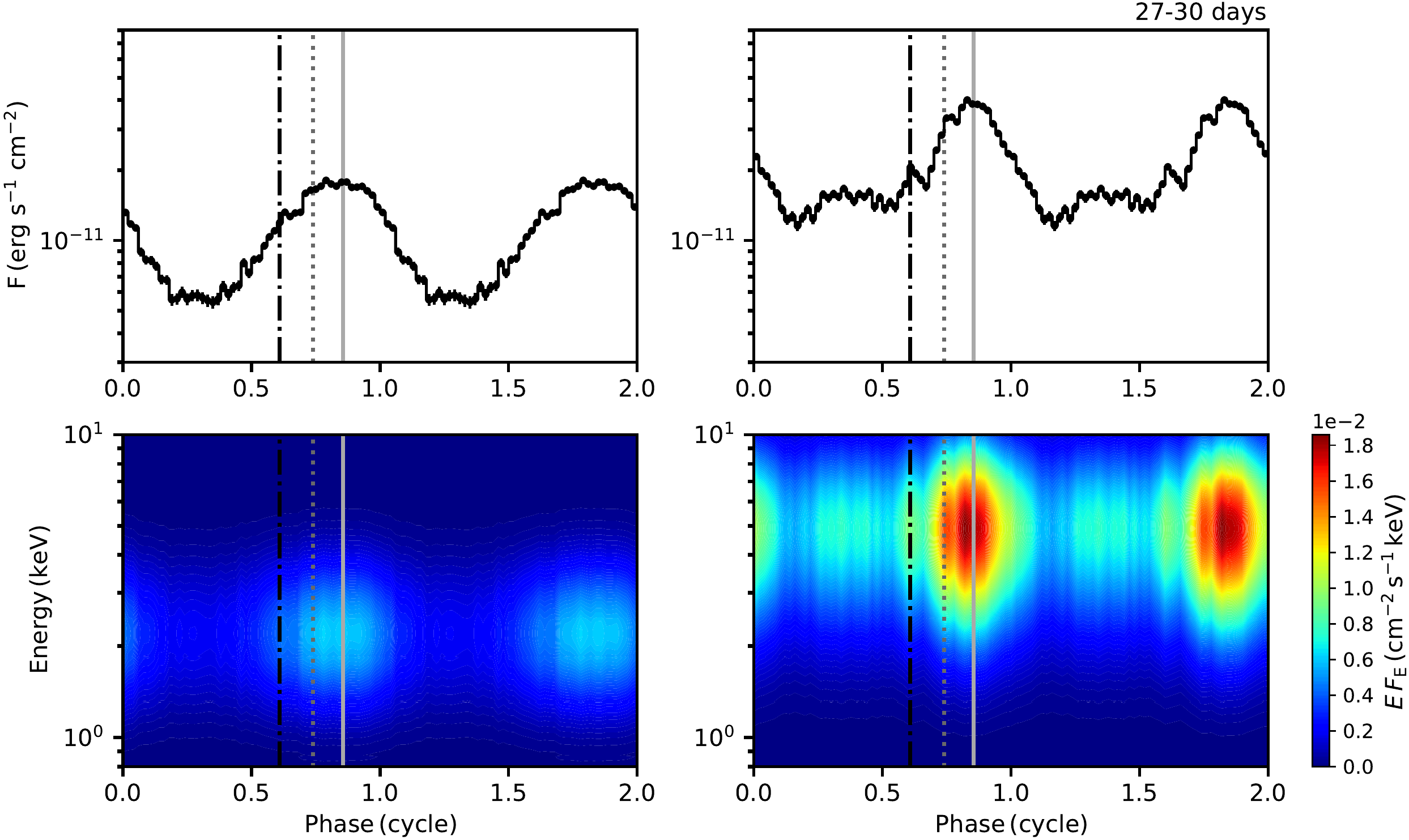}
  \includegraphics[angle=0,width=0.49\textwidth]{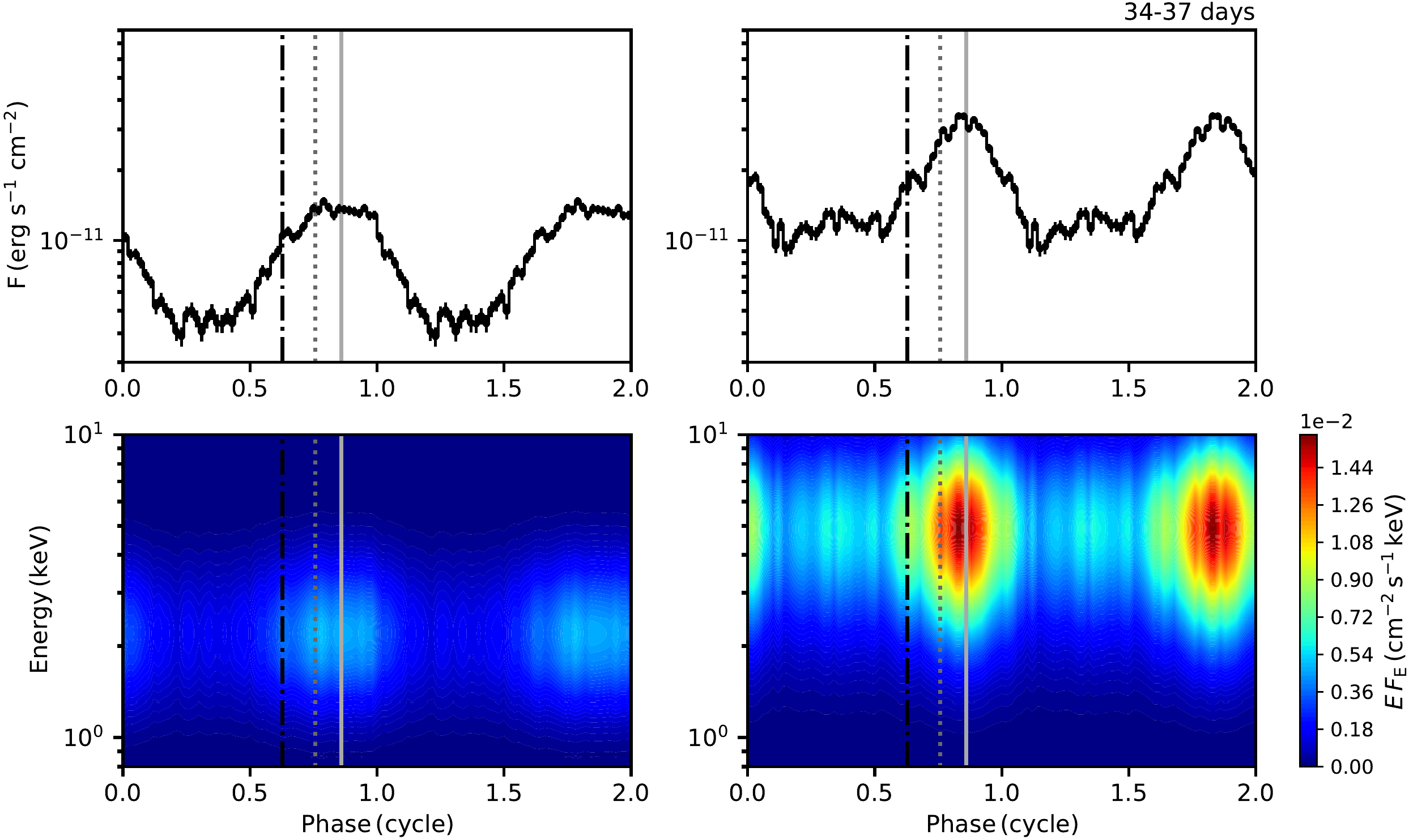}\\
  \includegraphics[angle=0,width=0.49\textwidth]{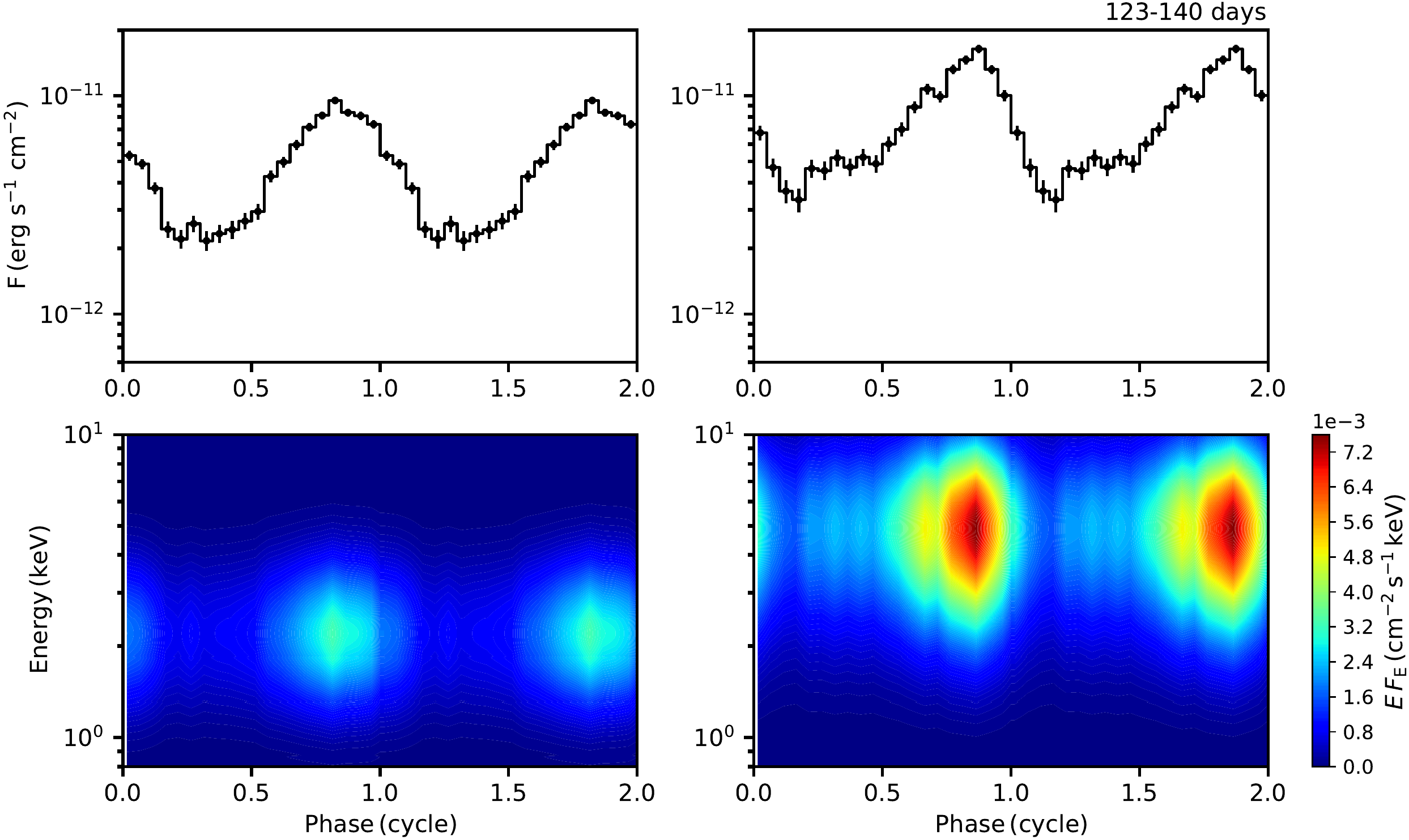}
  \includegraphics[angle=0,width=0.49\textwidth]{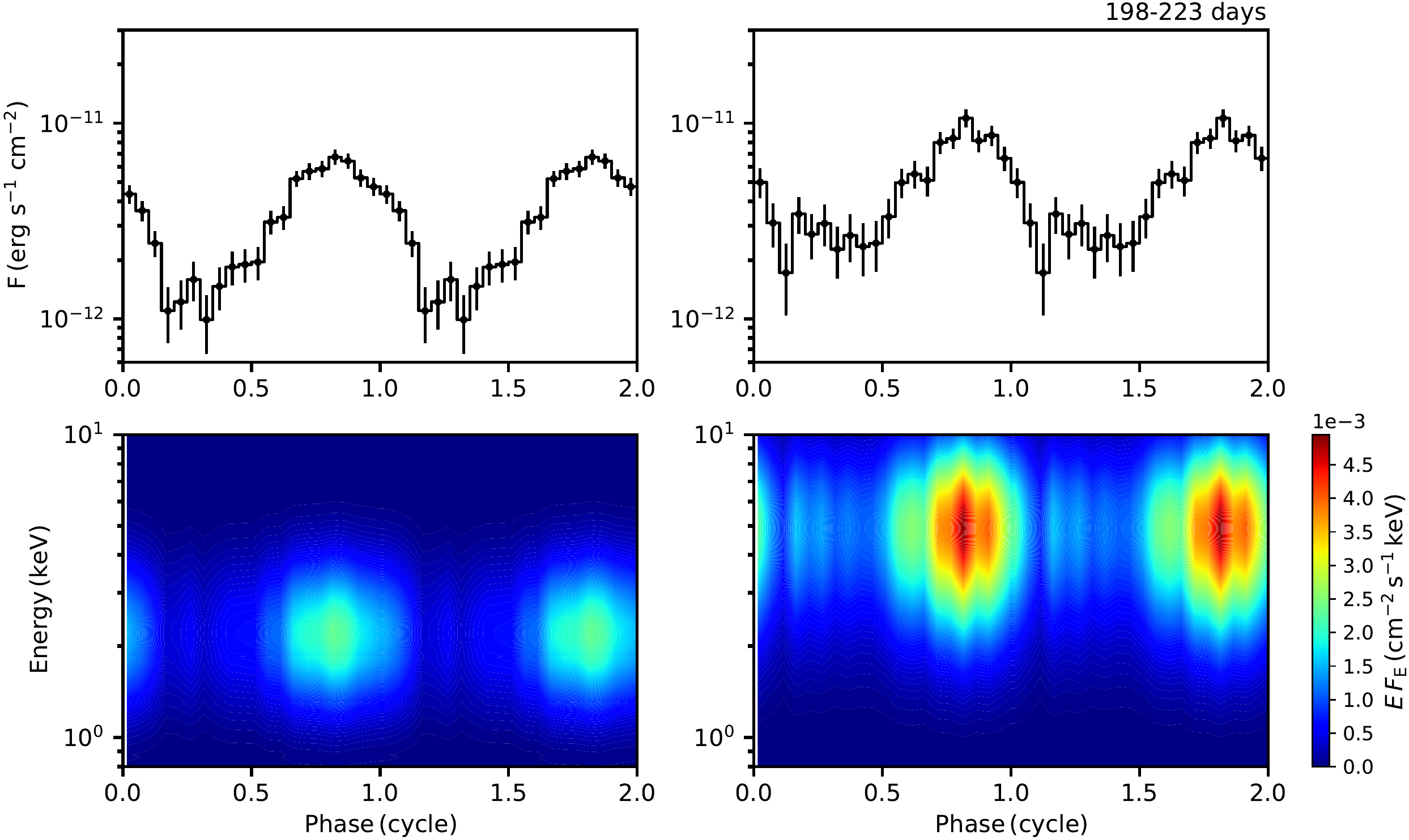}
\caption{Sample dynamic spectral profiles of \src\ derived by fitting 50 phase bins with an absorbed double-blackbody model. Panel-tetrads show the integrated fluxes (upper panels) and the $\nu F_{\rm \nu}$ photon flux contours in the phase-energy diagram (lower panels) for the warm BB (left) and hot BB (right) spectral components. The time interval for each is noted in the upper-right corner as days from outburst onset. The vertical lines indicate the best-fit Gaussian centroids to the count pulse peaks for each interval. The last row is for two intervals during the post-Sun-constrained period. The analysis is performed for 20 phase bins to increase S/N. See text for more details.}
\label{figDSP1}
\end{center}
\end{figure*}

Figure~\ref{figDSP1} shows a sample of our results and can be summarized as follows. Each panel-tetrad represents the epoch during which the analysis is performed. The upper panels show the 1--10~keV fluxes as a function of phase for the warm BB (left panels) and hot BB (right panels), respectively. The lower panels display the corresponding dynamic spectral profiles \citep[DSPs, e.g.,][]{rea09MNRAS:sgr0501} which show the photon flux (in units of photons~s$^{-1}$~cm$^{-2}$~keV) contours in phase-energy space. The integrated fluxes per phase-bin as well as the DSPs present a clear picture of the phase-variability pattern; the separate pulses in each profile are well resolved in both the warm and hot-BB components at the early stages of the outburst, though the trough between the peaks is more pronounced in the latter. The peak-separation becomes less evident with time, more quickly for the warm component. For instance, by days 18--21 post outburst, the peaks remain well resolved in the hot BB pulse profile while the warm BB profile had already simplified to a single-peak form. Interestingly, the pulse profiles of the last four epochs, 4 to 7 months later, appear stable, with a pure sinusoidal shape for the warm BB component, and a slightly more complex structure for the hot BB component, consisting of a double-peaked main pulse. Finally, the black dashed, dotted dark-gray, and solid light-gray lines in all panels prior to the Sun-constrained period represent the centroids of the Gaussian components that fit the three main peaks in the count pulse profiles as derived in Section~\ref{resPPE}; note their excellent agreement with the maxima of each peak as displayed in Figure~\ref{figDSP1}.

We plot the hot versus warm BB fluxes for each phase bin and each epoch in Figure~\ref{fig:hotvswarm}. These two parameters vary in tandem as a function of phase, maintaining a close flux ratio of $\sim2$. Additionally, this correlation seems to be non time-varying, with a similar correlation factor prevailing over the full length of the outburst epoch we are considering, including after the Sun-constrained period.

Finally, to check the time dependency of the three main pulse peaks in more detail, we extracted spectra centered on their respective Gaussian centroids and having a bin width $\Delta\phi=0.1$. We fit the time-dependent spectra of each peak simultaneously with an absorbed 2BB model. Initially, we only link the hydrogen column density, and track the evolution of the warm and hot BB temperatures. Even with much improved statistics, we find that the temperatures did not vary significantly with time for any of the peaks, and hence were linked together. The upper panels of Figure~\ref{fig:peakEvol} show the flux evolution of the warm (left) and hot (right) BB component for each peak. The straight lines are best-fit linear decay trends. We display the y-axis in log space, i.e., to represent the fractional change of the flux with time, evidently demonstrating that this quantity is constant between the three peaks and for both BB components. The lower panels are the corresponding change in area $R^2$ at a fiducial distance of 4~kpc.

\newpage

\section{Discussion}
\label{discuss}

\subsection{Summary of Results}

The \src\ near-daily \nicer\ monitoring campaign during the first 37 days of the outburst, enabled by the X-ray Timing Instrument's large effective area, afforded an in-depth look at the phase-resolved spectro-temporal and pulse-shape evolution of the source soft X-ray emission, among the most detailed such study of any magnetar to date.

\begin{figure}[t!]
    \centering
    \hspace{-0.2in}
    \includegraphics[width=0.47\textwidth]{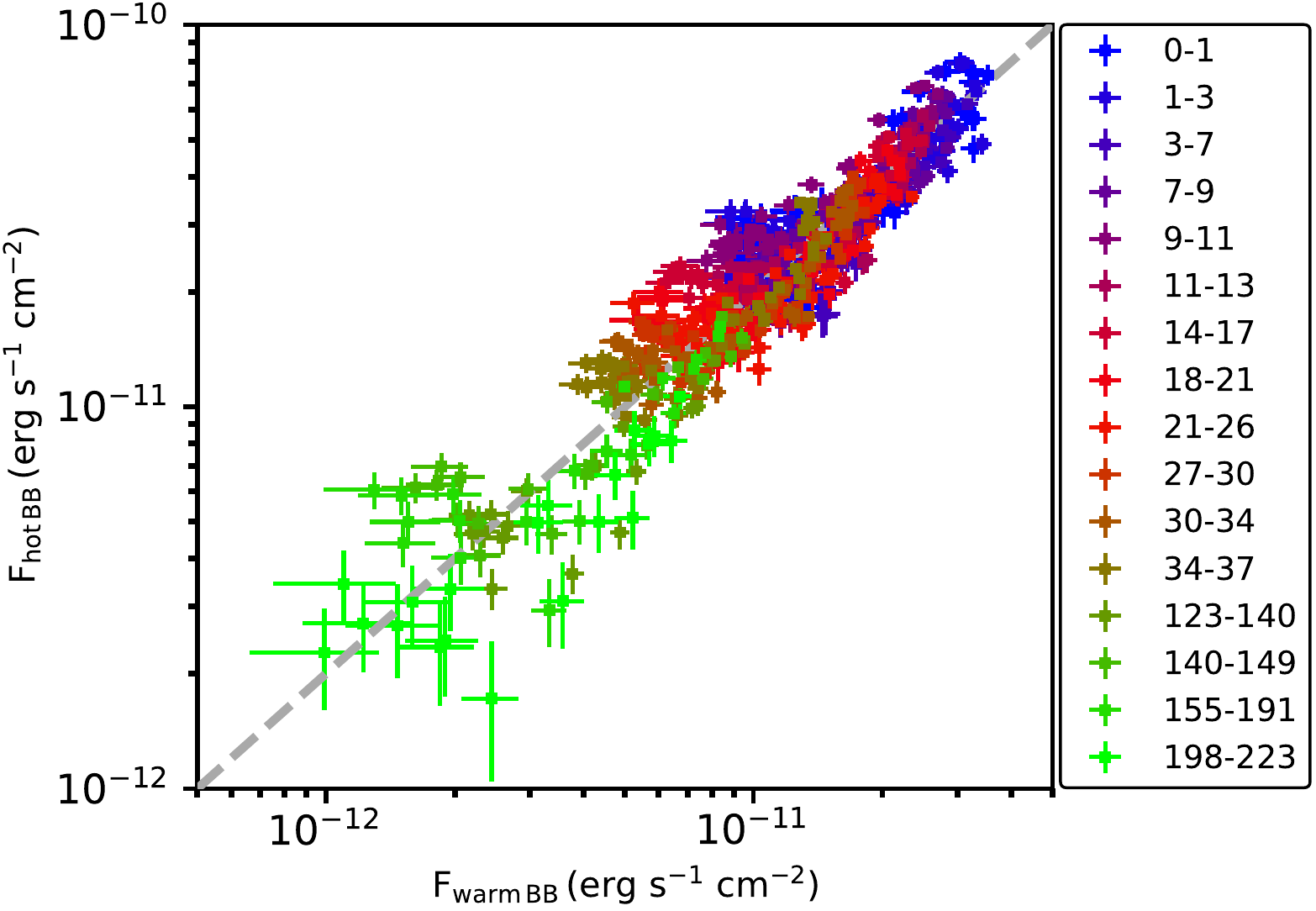}
    \caption{Hot BB vs warm BB fluxes derived from our phase-resolved spectroscopic analysis, color-coded by time interval (in days) during which the analysis is performed. An obvious linear correlation exists between the fluxes of each component throughout the outburst following $F_{\rm hot~BB} \propto 2 \times F_{\rm warm~BB}$, shown as a dashed gray line.}
    \label{fig:hotvswarm}
\end{figure}

Our fine (50 phase bins) phase-resolved spectroscopic analysis
indicates that two BBs, a warm and a hot component with temperatures
of about 0.5 and 1.2~keV, respectively, are necessary to fit all phase
bins. The BB temperatures show no apparent phase-variability nor do
they exhibit any significant evidence of change with time, despite the
complex shape and temporal evolution of the pulse profile, initially
marked with a triple peaked form \citep[see also][for phase-resolved
analysis with \xmm\ during the first two days of the
outburst]{cotizelati21ApJ1830}. Additionally, these two fluxes
maintain a close correlation throughout the rotational cycle of the
star, and show an almost identical decay trend with time, driven by
the linear decrease of the emitting surface area, as evidenced in
Figures~\ref{fig:hotvswarm} and \ref{fig:peakEvol}. Finally, the pulse
shapes of these thermally emitting regions are comparable; the peaks
appear at the same rotational phase, yet they are more pronounced for
the hotter BB component, which also possesses the smaller area among
the two (see Figure~\ref{figDSP1}).

The apparent shift in the \src\ thermal pulse peaks during the first
37 days of the outburst, in concert with the area shrinkage and flux
decay, is undoubtedly the most intriguing result of our
investigation. It is important to note that the motion is occurring in
the direction to significantly simplify the pulse shape from triple
peaked to almost single-pulsed, through the merging of the separate
peaks (see Figures~\ref{fig:ppCount} and \ref{figDSP1}). While this
aspect of pulse reduction during the decay phase of magnetar outbursts
is well documented in a few sources
\citep[e.g.,][]{rea13ApJ:0418,gavriil11ApJ:0142,scholz12ApJ:1822}, to
our knowledge, this is the first time that the process with which the
pulse shape simplifies is temporally and spectroscopically resolved.

\subsection{Nature of Pulse-peak Motion}

The nature of the pulse-peak motion is quite different from that
expected from most relevant physical processes of the star, thus
providing a strong constraint on its origin. For example, it is far
slower than the characteristic shear speed of the crust
$u_s=\sqrt{\mu/\rho}\approx 10^8\textrm{ cm s}^{-1}$, where $\mu,\rho$
are the shear modulus and mass density, respectively. Hot spot motion
is discussed in the context of accreting millisecond-pulsars
\citep[e.g.,][]{lamb09ApJ:hsm,patruno10ApJ:1807}, however, the same
mechanism cannot explain the behavior in \src\ considering the lack of
any sign of accretion, e.g., through spin-up (Y21). Heat conduction
might operate on a broadly similar timescale, but would not move local
hotspots around in a way that preserves their identity.

We envisage two physical scenarios that could be responsible for the
evolution of the persistent soft X-ray emission we observe during the
outburst: plastic motion of the crust, and untwisting of the
magnetosphere. These two scenarios are motivated by the most likely
origins of surface heating during magnetar outbursts; either from
energy deposition in the crust, e.g., due to Hall wave avalanches
\citep[][see also \citealt{beloborodov16:heat, Deibel17ApJ}]{li16ApJ:HW}, or bombardment of the
surface by accelerated particles in a twisted external B-field
\citep{beloborodov09ApJ, beloborodov16:heat}. In both cases the
outburst is initiated by an elastic failure of the crust, yet, their
evolution is dictated by different regions of the neutron star. These
two scenarios, which we consider in turn next, could act separately or
in concert to generate the peak merging and the hot spot area reduction.

\subsubsection{Plastic Motion of the Crust}

Given that the formation of the peaks and their subsequent motion
coincide with high-energy bursting activity aligning in phase with the
hot thermal regions (Y21), one logical scenario to
explore is whether we could be witnessing plastic motion of the crust
\citep{jones03,lyutikov15MNRAS}. Crustal motion could arise when
magnetic stresses gradually build up in the star's crust, eventually
exceeding its elastic yield limit and leading to a horizontal
displacement in a surface area \citep{thompson95MNRAS:GF,
  thompson00ApJ:1900,thompson02ApJ:magnetars}. Magnetic-field
evolution due to Hall drift for internal magnetic fields
$\gtrsim10^{14}$~G is a principal candidate for causing such large
stresses to develop within young NSs like magnetars \citep[][but see
\citealt{2017ApJ...841...54T} for an alternative
view]{goldreis92,cumming_AZ,gour16}. Within the model of
\citet{2016ApJ...824L..21L} (see also \citealt{2019MNRAS.486.4130L}),
the rate at which supra-yield stresses are converted into plastic
motion is set by the viscosity of the crust in its plastic phase
$\nu_{\rm pl}$, with the plastic-flow speed $u\sim l_{\rm char}
B^2/\nu_{\rm pl}$, where $l_{\rm char}$ is a characteristic
length scale associated with the flow. The pulse-peak motion of \src\
may, therefore, give us a probe of $\nu_{\rm pl}$ and the
poorly understood material properties of the neutron-star crust. 

\begin{figure}[t!]
\begin{center}
\hspace{-0.1in}
  \includegraphics[angle=0,width=0.23\textwidth]{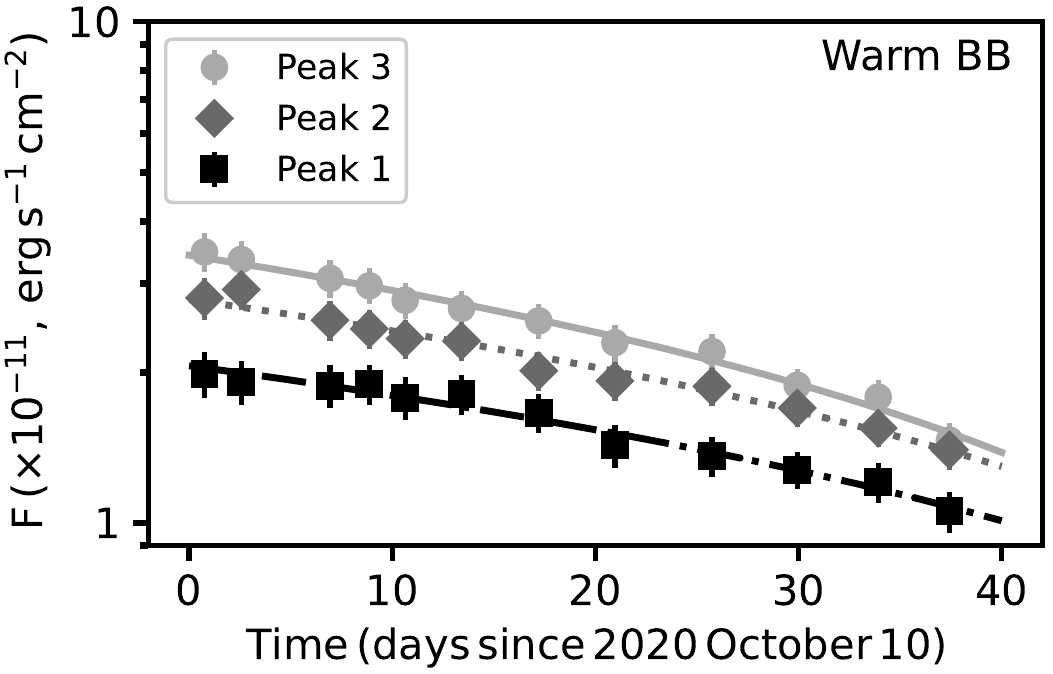}
  \includegraphics[angle=0,width=0.23\textwidth]{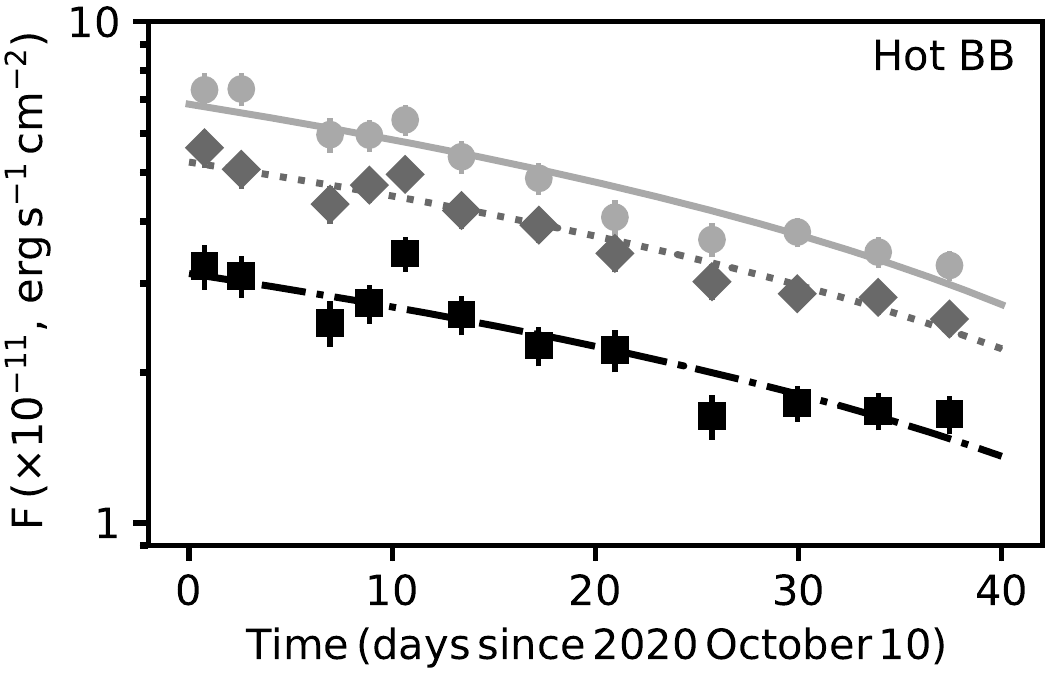}\\
\hspace{-0.1in}
  \includegraphics[angle=0,width=0.23\textwidth]{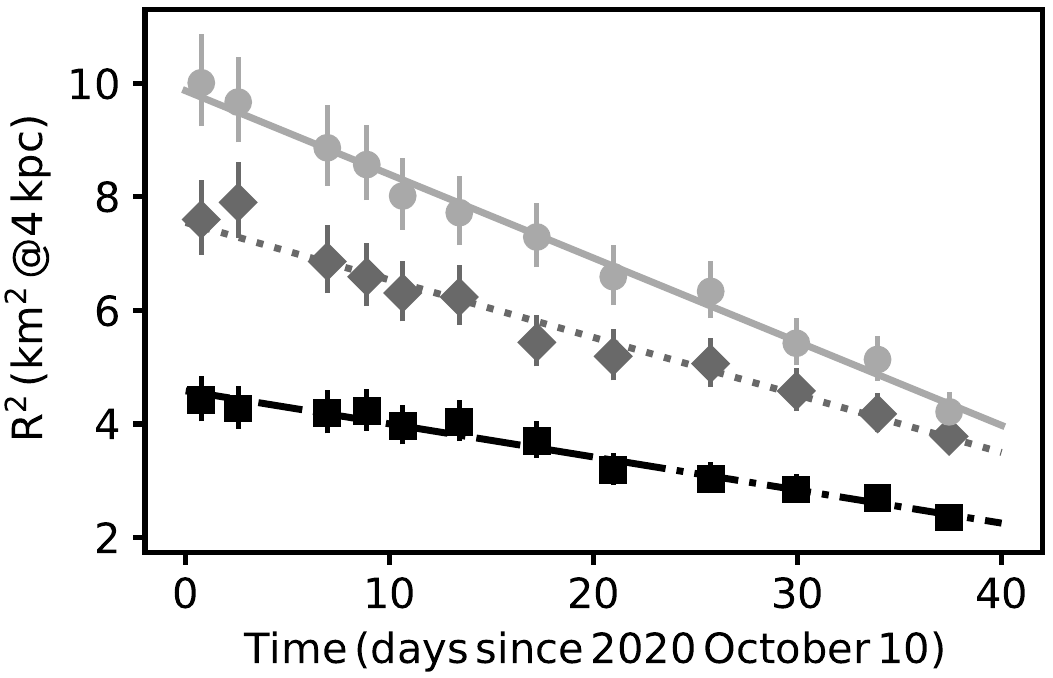}
  \includegraphics[angle=0,width=0.23\textwidth]{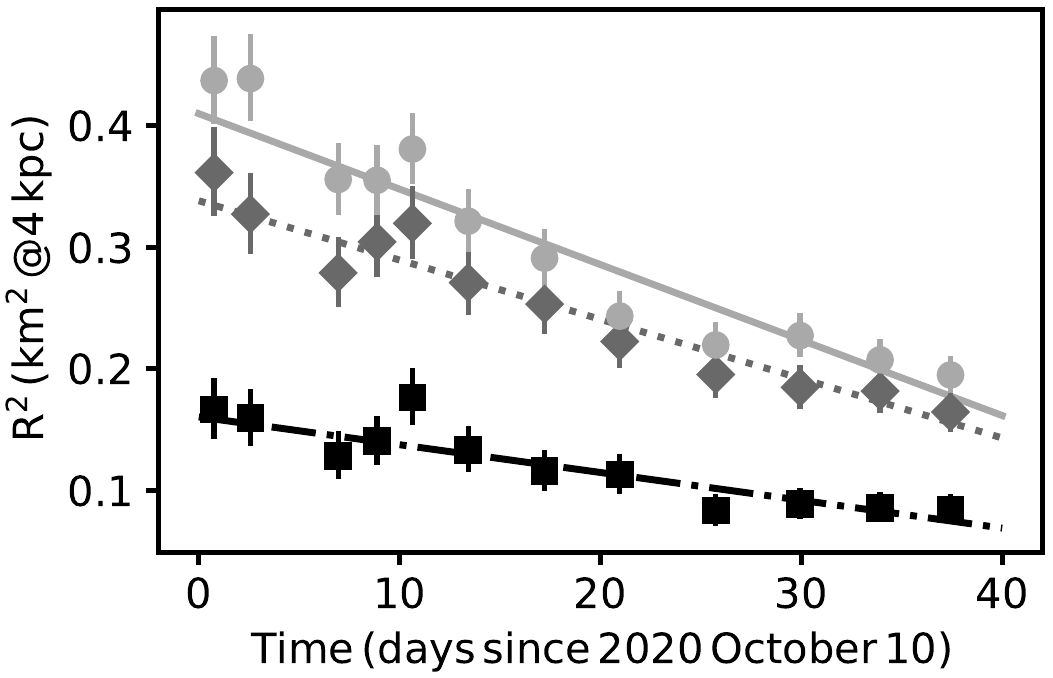}
\caption{Temporal evolution of the warm BB (left panels) and hot BB (right panels) for each of the three peaks in the \src\ pulse profile. The upper panels show the 1--10~keV flux decay while the lower panels display their corresponding area evolution. The lines are best-fit linear decay trends. Note that the vertical axes of the upper panels are logarithmic. The constant separations between the lines indicate a similar fractional change in each peak. See text for more details.}
\label{fig:peakEvol}
\end{center}
\end{figure}

Sustaining a plastic flow fast enough to power high-energy bursts
requires a rough balance between the effects of Hall drift and plastic
flow, corresponding to $\nu_{\rm pl}\sim 10^{36}-10^{38}\textrm{
  cm}^{-1}\textrm{g s}^{-1}$. However, this leads to a typical $u\sim
10-100\textrm{ cm yr}^{-1}$ in the inner crust
\citep{2019MNRAS.486.4130L,gourlan21}, roughly four orders of
magnitude slower than the $10^6$~cm~yr$^{-1}$ motion rate we
observe. Extrapolating from these results and using $u\propto
1/\nu_{\rm pl}$, the observed peak motion would need to be from a
region with $\nu_{\rm pl}\sim 10^{32}\textrm{ cm}^{-1}\textrm{g
s}^{-1}$, suggesting a location within the lower-density, more malleable
outer crust instead. 

The average BB temperature of roughly $1$ keV---equivalently
$T\approx10^7$ K---corresponds to the observed temperature of the
star's surface layers modified by the thin atmosphere. If the outburst
energy is deposited internally, the envelope below would act as an
efficient heat blanket, meaning that the outer crust will be
considerably hotter than the BB temperature: roughly $T\sim
10^8-10^9$~K for a typical crustal composition \citep[see,
e.g.,][]{kaminker14MNRAS}. This is above the melting temperature for
most of the outer crust \citep{HPY_book}; therefore in the emitting
region of \src, only the innermost part of the outer crust ($\rho
\gtrsim 10^{10} \textrm{ g cm}^{-3}$) is likely to be solid. 

In lieu of a quantitative calculation for $\nu_{\rm pl}$, we assume it
has the same density-dependent profile $\nu_{\rm pl}(\rho)$ as the
yield stress, but with a different prefactor \citep[see][for
details]{2019MNRAS.486.4130L} to estimate $\rho$ for the
region undergoing plastic motion. Assuming $T=10^9$~K below the
heat-insulating envelope and requiring that $\nu_{\rm pl}\sim
10^{32}\textrm{ cm}^{-1}\textrm{g s}^{-1}$ (in order to explain the
pulse-peak drift as plastic motion), we estimate $\rho\approx
10^{10}\textrm{ g cm}^{-3}$, corresponding to a depth of roughly $200$
metres---comparing with our previous estimate, this is also the
outermost region of the star that is likely to be solid. This is
important because the liquid regions of the outer crust cannot
sustain shear stresses, and so will experience motion on a fluid
timescale, far faster than the pulse drift we observe. We conclude
that, in the scenario of internal heating, the active region in \src,
responsible for both the X-ray bursts and thermal emission, should be the
near-molten inner part of the star's outer crust, undergoing plastic
flow.  Since less elastic energy can be stored here than in the inner
crust, this physical picture is also consistent with the X-ray bursts
being relatively weak (Y21). Interestingly, this
region of the crust is typically invoked to fit the outburst decay
trends with crustal thermal relaxation models \citep[e.g.,][]{
  rea2012ApJ:1822, scholz14ApJ:1822}.

\subsubsection{Untwisting the Magnetosphere}

The spectro-temporal evolution in \src\ can alternatively be explained
in connection with the paradigm of twisted magnetospheres. The
addition of toroidal components \teq{B_{\phi}} to generate twisted,
non-dipolar magnetar magnetospheres \citep{thompson02ApJ:magnetars}
occurs naturally with the presence of poloidal currents. A portion of
the energy involved in an impulsive dislocation process can be
deposited in the magnetosphere as field line footpoints promptly shift
across the stellar surface. Unfortunately, the lack of X-ray data
prior to the outburst onset in \src\ prohibits observation of
dislocation in action. Yet, the most constraining estimates for the
outburst rise time of $\lesssim2$~days, measured for several magnetars
\citep[][]{israel07ApJ:1647, esposito08MNRAS:1627,younes17:1935},
indicate that such rupturing is abrupt.

Once energy is stored in a twisted field configuration, it is
available for dissipation. Currents in twisted magnetospheres are
concentrated in restricted zones, current loops often called
$J$-bundles \citep{beloborodov09ApJ}, and in the axisymmetric ideal
MHD approximation, these assume quasi-toroidal shapes
\citep{chen17:mag}.  Therein, charges bombard the surface at the field
line footpoints and thereby generate hot spots/zones
\citep{beloborodov16:heat}. The pulse peaks observed for \src\
likely correspond to emission from such confined surface zones.

As the peaks migrate during the 37 day episode, the soft X-ray
spectrum does not vary, yet the effective area(s) of the hot and cool
surface emission regions monotonically decline according to
Fig.~\ref{fig:peakEvol}.  This behavior is commensurate with the
expectation \citep{beloborodov07ApJ:magCorona,beloborodov09ApJ} that
twisted fields slowly unwind and the volume of the current loops
declines due to Ohmic dissipation of the currents.  The nominal
timescale for this decay is of the order of a year
\citep{beloborodov09ApJ}, albeit determined for global
  axisymmetric twisted configurations. The phase separation of the
peaks at the outset of the migration likely corresponds to a
separation in the range of \teq{20-50^{\circ}} across the surface.
This indicates that the twisted zones are flux tubes/ropes akin to
those observed in the solar corona,\footnote{Images of solar
  prominences and coronal field loops can be found in the Solar
  Dynamics Observatory gallery at {\tt
    https://www.nasa.gov/mission\_pages/sdo/images/index.html}} as
opposed to extended quasi-toroidal volumes. The smaller area of the
hotter blackbody component naturally suggests an umbral/penumbral
configuration for the twisted field region as it threads the stellar
surface, likely persisting along the tube to its apex: the magnetic
twist energy is \teq{\propto (B_{\phi})^2} and is greater at the
center of the tube than at its periphery. The shrinkage of the hot
spots, concurrent with the peak-migration observed for \src\ may
correspond to shrinking a flux tube on the $\sim 1$-month timescale, a
dynamical corona event that has not been discerned before for
magnetars.

This field untwisting is accompanied by a migration of current
bombardment regions across the surface \citep{beloborodov09ApJ,
  chen17:mag}, also causing a modest change in direction of the field
at the footpoints of the tube's field lines. The surface penetration depth 
of the bombarding pairs' relativistic electrons is no more than a few cm due 
to the large optical depth in the surface magnetized atmosphere  
(e.g., \citealt{ho07MNRAS:1856,  gonzalez19MNRAS}).   Accordingly, 
the heating signatures of such bombardment will be intimately tied 
to the soft X-ray atmospheric emission. A strong local radiation anisotropy is expected from the surface, with a
preferential beaming along B of soft X rays at frequencies well
below the cyclotron fundamental $eB/m_ec$
\citep{vanAdelsberg06MNRAS,taverna20MNRAS,Barchas-2021-MNRAS}. Hence, the pulse-peak migration, in this
scenario, would be a combination of physical crustal motion as the
external field-lines untwist, and a non-negligible contribution from
changes in the footpoint B direction. Here, the crustal motion
is likely experienced at a depth where the activated field lines
anchor to the rigid surface; a poorly known quantity in the case of
non-dipolar field lines.

During the untwisting, the energy reservoir of the excess magnetic
field is transferred to charges and radiation. The charges
(\teq{e^{\pm}} pairs) are energized by dynamical electric fields in
the magnetosphere and in principle can generate hard X-ray signals via
the resonant inverse Compton scattering of surface X-ray photons
\citep[e.g.,][]{baring07,beloborodov13ApJ,wadiasingh18ApJ}. Such
an emission component is not dominant for \src\
\citep{cotizelati21ApJ1830}, indicating that the twist energy is
ultimately deposited through bombardment in heating the surface. Thus
the soft X-ray luminosity provides an observational upper limit to the
untwisting field decay rate.

The twisted flux tube can be ascribed an effective volume 
\teq{V_{\rm ft} = \lambda \rns A_{\rm ft}}, where \teq{A_{\rm ft}} is the footprint 
area of the tube at the stellar surface, and \teq{\rns\sim 10}km is the neutron star radius.
Also, \teq{\lambda} describes the length along the tube for the dominant contribution 
to the magnetic energy \teq{U_{\rm tw}} stored in the twisted field configuration 
{\it in excess} of the dipole value.  Nominally, \teq{\lambda \sim 1-10}, with poloidal 
\teq{B_r} and \teq{B_{\theta}} components contributing significantly
\citep{thompson02ApJ:magnetars,Pavan-2009-MNRAS} to this excess. 
The baseline energy scale for the field is the total dipolar contribution 
within the flux tube (ft) at the conclusion of the peak migration, which is
\teq{U_{\rm d} = (B_p^2/8\pi)  \, \rns A_{\rm ft} \approx  1.1\times 10^{44}\, 
[ A_{\rm ft}/ 1\, {\hbox{km}}^2]}erg for a surface polar field of 
\teq{B_p=5.4 \times 10^{14}}G.  Using this, one can then express the energy 
of the twisted field reservoir, \teq{U_{\rm tw}}, and bound it observationally 
by the cumulative energy radiated in the soft X-rays during the 37 day migration epoch:
\begin{eqnarray}
   U_{\rm tw} & \;\sim\; &  1.1\times 10^{44}\, f\lambda\, \left( \frac{B_{\phi}}{B_p} \right)^2 \, 
   \frac{\vert \Delta A_{\rm ft}\vert }{1\, {\hbox{km}}^2} \; \hbox{erg} \nonumber\\[-5.5pt]
 \label{eq:U_tw_def}\\[-5.5pt]
   && \;\lesssim\; \Delta E_X \; \sim\; 3.2 \times 10^{41}\, \hbox{erg}\quad .\nonumber
\end{eqnarray}
Here \teq{f\sim 1-10} is a factor that accounts for the added contribution of the poloidal 
components to the twisted field.   The cumulative radiated energy \teq{\Delta E_X} 
assumes the luminosity \teq{L_X\sim 10^{35}}erg/sec for a source distance of 4 kpc,
and is herein ascribed to the umbral hot component, for which 
\teq{-\Delta A_{\rm fp}\sim 0.1-0.3}km$^2$ from Fig.~\ref{fig:peakEvol}.  For 
\teq{f\lambda \sim 1}, this inequality constrains the toroidal field at the outset of the migration 
to \teq{B_{\phi}/B_p\lesssim 0.1}, which is in keeping with theoretical expectations for 
moderately strong twists \citep{thompson02ApJ:magnetars,chen17:mag}.
Given that \teq{f\lambda} is likely larger, around 10--30, 
the twist is correspondingly smaller.  Accordingly, forging a connection of twisted fields 
to migrating hotspots enables important constraints on the twist morphology in \src.

\subsection{Concluding Remarks}

Many reasons could have precluded us from observing this behavior
during previous magnetar outbursts. For instance, daily
high-throughput X-ray observations were extremely rare prior to the
launch of \nicer\ and while only a few (e.g., \xmm) observations may
have been able to spot the peak motion we report here, typically these
are spread over the entire outburst decay period of months to years,
insufficiently sampling such comparatively rapid
evolution. Nonetheless, \nicer\ has performed regular daily monitoring
observations of a few more transient magnetars, Swift J1818.0$-$1607
\citep{hu20ApJ:1818}, and Swift J1555.2$-$5402
\citep{enoto21:1555}. Both these magnetars showed a very stable
single-pulsed profile over the first months of the outburst (note,
however, the lower number counts compared to \src\ due to the
larger absorbing hydrogen column density in their direction). This may
imply that scarcity of observations alone cannot account for the
absence of pulse-peak motion, and some intrinsic properties of the
outburst must be in play---e.g., the geometry of the affected regions
and/or the depth at which crustal motion is occurring. Continued daily
monitoring of magnetar outbursts with \nicer\ is therefore critical to
reveal the diversity of this phenomenon among the population. This in
turn could guide the development of a more complete theoretical
picture where magnetosphere and crust are considered in tandem, e.g.,
by defining how the magnetic field threads from the magnetosphere
through the atmosphere to the outer crust. Such investigations will
elucidate the crustal and magnetospheric physics, helping to
discriminate between these scenarios as the origin for pulse peak
migration.

\section*{Acknowledgments}

G.Y.'s research is supported by an appointment to the NASA Postdoctoral Program at the Goddard Space Flight Center, administered by Universities Space Research Association under contract with NASA. Prior to the appointment, G.Y. was supported by funding through the NASA NICER Guest Observer program grant 80NSSC21K0233 and the NASA Fermi Guest Investigator program grant 19-19FermiC13-0021. M.G.B. acknowledges the generous support of the National Science Foundation through grant AST-1813649. ZW is partly supported by NASA under award number 80GSFC21M0002 as well as by an appointment to the NASA Postdoctoral Program at the Goddard Space Flight Center, administered by Universities Space Research Association under contract with NASA. W.C.G.H. acknowledges support through grant 80NSSC20K0278 from NASA.

\end{document}